# Semi-classical electronic transport properties of ternary compound AlGaAs$_2$: Role of different scattering mechanisms.

Soubhik Chakrabarty[a], Anup Kumar Mandia[b], Bhaskaran Muralidharan[b], Seung Cheol Lee[a,*], and Satadeep Bhattacharjee[a,*]

[a]Indo-Korea Science and Technology Center, Bangalore 560065, India

[b]Indian Institute of Technology, Mumbai-400076, India

*Corresponding Author E-mail: seungcheol.lee@ikst.res.in,
satadeep.bhattacharjee@ikst.res.in

**Abstract:**

We present a comprehensive investigation of semi-classical transport properties of *n*-type ternary compound AlGaAs$_2$, using Rode's iterative method. Four scattering mechanisms, have been included in our transport calculation, namely, ionized impurity, piezoelectric, acoustic deformation and polar optical phonon (POP). The scattering rates have been calculated in terms of *ab-initio* parameters. We consider AlGaAs$_2$ to have two distinct crystal geometries, one in tetragonal phase (space group: $p\bar{4}m2$), while the other one having body centered tetragonal crystal structure (space group: $I\bar{4}2d$). We have observed higher electron mobility in the body centered tetragonal phase, thereby making it more suitable for high mobility device application, over the tetragonal phase. In order to understand the differences in electron moblities for these two phases, curvatures of the E-k graph of the conduction bands for these phases have been compared. At room temperature, the dominant contribution in electron mobility was found to be provided by inelastic POP scattering. We have also noted that mobility is underestimated in relaxation time approximation as compared with the Rode's iterative approach.

## I. Introduction

III-V compound semiconductors such as GaAs, GaSb, AlAs, InP etc. possessing direct band gap have drawn enormous attention of the scientific community over the past few decades because of their strong potential for application in high mobility electronic and high-performance optoelectronic devices [1-9]. However the binary III-V compounds lack the flexibility of tuning theirs band structure which imposes limitations to some extent on their device applications. This limitation has brought the spotlight on the research and development of III-V ternary and quaternary systems [1-5]. In order to exploit the application potential of a material in device design it is necessary to have a precise knowledge of their electronic structure (band dispersion, band gap, density of states (DOS) etc.) and electron transport properties (estimation and electron mobility and its variation with temperature and carrier concentration etc.)

In an ideal, perfect periodic crystal there is no scattering of electrons. However, in real crystal, electrons are scattered due to lattice vibrations, presence of impurities, dislocations etc. which determines the electron distribution function. We need to solve Boltzmann transport equation (BTE) to obtain the electron distribution function which governs the electron transport properties viz. mobility, conductivity etc. Thus modeling of semi-classical electron transport through semiconductors hinges on the numerical solution of BTE [10-25].

Most of the available models for solving BTE employ relaxation time approximation [12,13,16-22]. There are some semi-empirical models that considers input parameters viz. effective mass, band gap, polar optical phonon frequency, dielectric constant from experimental data for calculating the scattering rates [10-15]. In these type of models parabolic or Kane energy-momentum dispersion relation is considered. The relaxation time is assumed to follow a power law distribution of energy. However this basic assumption fails for inelastic scattering for which scattering rate does not satisfy such power law dependency on energy [10,15,20]. Moreover these models rely on the availability of experimental data, thereby limiting the predicting ability of these models for new materials. There are some other RTA models that relies on *ab-initio* band structure [26,27]. The basic assumptions of these models are consideration of the electron-phonon scattering to be elastic, the distribution function to be unchanged from its equilibrium form and relaxation time to be a constant (c-RTA). Madsen and Singh [27] have witnessed that this c-RTA models works goods for materials having scattering rate moderately constant. However the over simplified assumptions of these models ruins the predictive power of these approaches.

The RTA models are suitable when the scattering mechanisms are elastic and isotropic, and as a consequence relaxation time becomes independent of the distribution function. Polar optical phonon (POP) scattering have significant influence on the distribution function at room temperature for polar III-V compounds [9,28,29]. In III-V compounds oscillating electric dipole is generated because of the movement of charged ions in the unit cell and the corresponding vibrational mode is known as the polar optical phonon mode. The interaction of the conduction electron with POP is inelastic and nonrandomizing, making RTA inappropriate [10,15,20,22] for describing transport phenomena of the III-V materials at room temperature. Rode's iterative method [10,15,20-25] is an effective solution for the incorporation of POP scattering in order to simulate semi-classical transport phenomena of III-V materials.

In the present work we have calculated the mobility of $n$-type ternary compound AlGaAs$_2$, with Al:Ga:As ratio being 1:1:2 using Rode's iterative method. The input parameters viz. band dispersion, DOS, dielectric constant, deformation potential, POP frequency, wave function admixture, required for calculating different scattering rates have been calculated using density functional based approach in which the only input is the crystal geometry. In our previous work [30] we computed the mobility of $n$-type ZnSe using Rode-*ab-initio* approach and observed good agreement with the experimental results. In this present work we have considered ternary AlGaAs$_2$ compounds having two distinct crystal structures. This paper aims to provide a comparison of electron mobility of the two ternary compounds having different atomic arrangement with same stoichiometry and predict the better one for high-speed electronic devices on the basis of electron mobility.

## II. Methodology

### A. Solution of Boltzmann transport equation:

Semi-classical transport calculations have been performed using our code AMMCR [31]. Brief methodology of solving the Boltzmann transport equation (BTE) is presented below.

Under the application of a low electric filed *E*, BTE for the electron distribution function *f* is given by

$$\frac{df}{dt} + \mathbf{v}(k).\nabla_r f + \frac{eE}{\hbar}.\nabla_k f = \left(\frac{\partial f}{\partial t}\right)_S \tag{1}$$

where $\mathbf{v}(k)$ represents group velocity of electron and $\left(\frac{\partial f}{\partial t}\right)_S$ denotes change in the distribution function due to all scattering processes.

Under steady state condition $\frac{df}{dt} = 0$, and under the absence of thermal driving force ( spatial homogeneity) the second term in equation (1) vanishes. Under these two conditions equation (1) can be rewritten as

$$\frac{eE}{\hbar}.\nabla_k f = \int \{s(k',k)f(k')(1-f(k)) - s(k,k')f(k)(1-f(k'))\}dk' \tag{2}$$

where $s(k',k)$ represents scattering rate for an electron making a transition from a state $k$ to a state $k'$.

Due to the application of low electric filed the distribution function is assumed (linear response) to get perturbed as follows

$$f(k) = f_0[\varepsilon(k)] + x g(k) \tag{3}$$

where $f$ is the actual perturbed distribution function, $f_0$ represents the equilibrium part of the distribution given by the Fermi-Dirac distribution function, $g(k)$ is the perturbation part caused by the application of low electric field and $x$ denotes the cosine of the angle between $k$ and the electric filed. In order to calculate low-field electron transport properties we need to calculate the perturbation $g(k)$. After performing some mathematical steps and assuming $x = 1$, BTE can be expressed to yield $g(k)$ as follows

$$g_{i+1}(k) = \frac{S_i(g_i(k)) - \frac{eE}{\hbar}\frac{\partial f}{\partial k}}{S_0(k) + \frac{1}{\tau_{el}(k)}} \tag{4}$$

The scattering rates in equation (2) has two components; elastic part ($s_{el}$) and the inelastic part ($s_{in}$). i.e. $s(k,k') = s(k,k')_{el} + s(k,k')_{in}$.

$S_i$, $S_0$, $\tau_{el}$ appearing in equation (4) are given by

$$S_i(g_i(k)) = \int X g_i(k')\left[s_{in}(k',k)(1-f(k)) + s_{in}(k,k')f(k)\right]dk' \tag{5}$$

$$S_0(k) = \int \left[s_{in}(k,k')(1-f(k')) + s_{in}(k',k)f(k')\right]dk' \tag{6}$$

$$\frac{1}{\tau_{el}} = \int (1-X) s_{el}(k,k') dk' \quad (7)$$

where *X* in equation (7) corresponds to the cosine of the angle between final and initial wave vectors.

$S_i$ and *f* are functions of *g*, hence equation (4) has to be solved iteratively using Rode's iterative method in order to get the converged value of $g(k)$. The two term $S_i$, $S_0$ deal with the inelastic POP scattering and the tem $\tau_{el}$ captures the effect of all the elastic scattering processes. According to Matthiessen's rule the total elastic scattering rate $\frac{1}{\tau_{el}(k)}$ can be written as the sum of the momentum relaxation rates off all the scattering processes.

$$\frac{1}{\tau_{el}(k)} = \frac{1}{\tau_{ii}(k)} + \frac{1}{\tau_{pz}(k)} + \frac{1}{\tau_{ac}(k)} \quad (8)$$

where the subscripts *ii*, *pz*, *ac* respectively corresponds to the ionized impurity, piezoelectric, acoustic deformation potential scattering processes.

The rates of the different elastic scattering processes have been calculated in terms of electron group velocity and density of states as discussed in the literature [30-32]. Inelastic POP scattering rate has been calculated iteratively using Rode's iterative formalism details of which is presented in the previously published reports [30-32]. In our previous work [30] we have followed identical formalism for calculating low-filed transport properties of *n*-ZnSe.

The carrier mobility has been computed using the following expression

$$\mu = \frac{1}{3E} \frac{\int v(\varepsilon) D_s(\varepsilon) g(\varepsilon) d\varepsilon}{\int D_s(\varepsilon) f(\varepsilon) d\varepsilon} \quad (9)$$

where $D_s(\varepsilon)$ is the density of states. The group velocity of electron is calculated from *ab-initio* band dispersion of the conduction band by using the following expression

$$v(k) = \frac{1}{\hbar} \frac{d\varepsilon}{dk} \quad (10)$$

**B. *Ab-initio* inputs:**

Band structure and density of states of the ternary AlGaAs$_2$ compounds have been computed using density functional theory as implemented in Vienna *ab-initio* simulation package (VASP) [33-35]. In order to calculate carrier velocities we have calculated the band structures using a highly dense k mesh around the conduction band minimum (CBM) and then we expressed the average electron energies for the conduction band as a function of distance

$k = |\mathbf{k}|$ from the CBM. After performing the numerical fitting of the conduction band with a six degree polynomial we computed carrier group velocity using equation (10). This approach produce a smooth curve for mobility and has been reported earlier by Alireja *et.al* [32]. For the carrier concentration (*n*) we have considered the following equation

$$n = \frac{1}{V_0} \int_{\varepsilon_c}^{+\infty} D_s(\varepsilon) f(\varepsilon) d\varepsilon \qquad (11)$$

where $V_0$ is the volume of the relaxed unit cell. Fermi level for a given carrier concatenation is computed by matching the concentration according to equation (11).

For calculating deformation potential ($E_D$) we calculated the changes in the CBM by changing the volume of the unit cell and calculated $E_D$ using the following expression

$$E_D = -V \left( \frac{\partial E}{\partial V} \right) \bigg|_{V=V_0} \qquad (12)$$

We used density functional perturbation theory [36,37] for calculating piezoelectric constants, low and high frequency dielectric constants, frequency of polar optical phonons as impediment in the VASP code. For obtaining elastic tensor, finite distortions of the lattice was considered and the elastic constants was derived from the stress-strain relationship [38]. The elastic tensor has been computed for both, fixed atoms, as well as performing relaxation of them. After obtaining the elastic matrix from VASP output, we used MechElastic [39] script to obtain the longitudinal and transverse elastic constants.

### C. *Ab-initio* computational details:

We have considered ternary AlGaAs$_2$ compounds of two different crystal geometries. One of the configurations crystallizes in tetragonal (TET) lattice with space group type $p\bar{4}m2$ (115). The other one crystallizes in body centered tetragonal (BCT) lattice with space group type $I\bar{4}2d$ (122). We obtained the ternary structures through cation substitution in a GaAs supercell, using Site-Occupation Disorder package [40]. For geometry optimization and electronic structure calculation we have used DFT based approach as implemented in VASP code. We considered generalized gradient approximation (GGA) of Perdew, Burke, and Ernzerhof (PBE) [41] to approximate the exchange-correlation part. In order to to describe the electron-ion interaction we have employed projector augmented wave (PAW) method [42]. We used conjugate-gradient method [43] for ionic relaxation. The Hellman-Feynman forces on the constituent atoms were minimized with the tolerance of 0.005 eV/Å. We considered 23×23×15 and 13×13×13 Monkhorst-Pack [44] k-mesh for sampling the Brillouin Zones (BZ) of the TET and BCT configurations, respectively. In order to calculate group velocity of the electrons in the conduction band we performed band structure calculation with a high dense k-mesh around the CBM. In order to obtain phonon dispersion, we considered finite displacements of atoms in a 3×3×3 supercell and the force sets were obtained using Phonopy [45] code from VASP output.

III. **Results and Discussion:**

A. **Electronic structure:**

16 atom unit cell with Al:Ga:As ratio being 1:1:2 for the two phases is shown in figure 1. We consider primitive unit cells (figure S1 of supplementary material) for electronic structure calculations. The primitive unit cells were obtained by imposing symmetry on the 16 atom unit cells as implemented in Phonopy code. The primitive unit cells of the TET and BCT phases consist of 4 and 8 atoms respectively. Optimized lattice parameters of the primitive cell and the angles between the primitive translation vectors are given in table I. PBE estimated band structure for both the TET and BCT configurations predicts semiconducting nature. Both the configurations are found out to be direct band gap ($E_g$) semiconductors with valence band maximum and conduction band minimum situated at the BZ centre (figure 2). PBE estimated band gap values for the TET and BCT configurations are found out to be 0.99 eV and 0.86 eV. We have analyzed atom and orbital projected DOS (figure S2) in order to investigate the contribution of different atoms and orbitals on the valence and conduction bands. For both the TET and BCT configurations we observed that As-p states has the dominant contribution to the valence band. However, the major contribution in the conduction band is coming from Ga-s and As-p states.

B. **Stability:**

In order to analyze dynamical stability of the two configurations we have plotted the phonon dispersion in figure 3. No imaginary frequency has been observed for the BCT phase. However, for the TET phase imaginary frequency of magnitude less than 4 cm$^{-1}$ has been observed around the BZ centre. This small imaginary acoustic phonon appearing near the zone center do not correspond structural instability. The negligible imaginary frequency of the acoustic mode is a numerical error, arising due to violation of transnational invariance in approximated calculation [46,47]. The phonon dispersion plots, hence confirms the structural stability of the both the phases.

We have studied mechanical stability of the two phases using Born stability criteria [48]. The necessary stability criteria for tetragonal systems are given by

(i) $C_{11} - C_{12} > 0$

(ii) $2C_{13}^2 < C_{33}(C_{11} + C_{12})$

(iii) $C_{44} > 0$

(iv) $C_{66} > 0$

(v) $2C_{16}^2 < C_{66}(C_{11} - C_{12})$

Coefficients of the elastic matrix $C_{ij}$ obtained with DFT-PBE calculations for both phases have been found to satisfy all the above mentioned criteria, thereby suggesting their mechanical stability.

### C. Transport properties:

**Mobility vs. temperature:** The variation of mobility vs. temperature for different carrier concentration is shown in figure 4 for the BCT configuration. Mobility continuously decreases with temperature as expected. Mobility values do not differ much for low carrier concentration viz. $1\times10^{13}$ cm$^{-3}$ and $1\times10^{15}$ cm$^{-3}$. This is because of the fact that at low carrier concentration ionized impurity is less significant. Figure 4b shows the comparison of mobility estimated using RTA method and Rode's scheme. In RTA approach the mobility is underestimated. This is attributed to the fact that, POP scattering is inelastic and nonrandomizing and hence the perturbation in the distribution function using relaxation time ( either constant or power law dependency on energy) cannot be defined. At low temperature POP scattering become insignificant, as a result of which mobility estimated using RTA and Rode's iterative method become almost equal. The variation of mobility with temperature for the TET is given in figure S3 which shows similar trend as that of the BCT phase. Comparing the mobility for the TET and BCT configurations at a carrier concentration of $1\times10^{17}$ cm$^{-3}$ (figure S4) we observed that the BCT configuration shows higher mobility as compared with the TET configuration for the entire range of temperature. In figure 5 we have plotted the average energy of electron for the conduction band vs. the k-distance from the CBM for both the configurations. The curvature of the E-k graph for the BCT is higher as compared with the TET, and this attributes to the higher mobility of the BCT structure.

**Mobility vs. carrier concentration:** In figure 6 we have plotted the variation of mobility with doping concentration $(n)$ at 50K and 300K temperature for the BCT configuration. We have observed that for both of the configurations mobility does not change significantly for $1\times10^{10} \leq n \leq 1\times10^{13}$ at T=50K and $1\times10^{10} \leq n \leq 1\times10^{15}$ at T=300K. At T=50K/300K when $n$ is increased beyond $10^{13}/10^{15}$ cm$^{-3}$ mobility starts decreasing. In order to gain an insight we have analyzed the variation of different components of mobility with doping concentration. According to Matthiessen's rule

$$\frac{1}{\mu} = \frac{1}{\mu_{ii}} + \frac{1}{\mu_{po}} + \frac{1}{\mu_{ac}} + \frac{1}{\mu_{pz}} \tag{13}$$

where $\mu$ is the total mobility and the suffixes ii, po, ac, pz are used to indicate ionized impurity, POP, acoustic deformation potential, piezoelectric scattering mechanism. $\mu_{ii}$ is the mobility of the material considering only the ionized impurity scattering mechanism; $\mu_{po}$ is the mobility, if the scattering occurs only through POP scattering mechanism and so on. $\mu_{po}$, $\mu_{ac}$ and $\mu_{pz}$ almost remains constant for the entire range of doping concentration where as $\mu_{ii}$ strongly depends on $n$, as is evident from figure 7 for the BCT configuration. As the different components appears in reciprocals in equation (13), the component showing the smallest value is the most significant one. At T=50K, the dominant contributions comes from $\mu_{ac}, \mu_{pz}$ for $n \leq 1\times10^{13}$ cm$^{-3}$. For $n \geq 1\times10^{13}$ cm$^{-3}$, $\mu_{ii}$ becomes comparable with $\mu_{ac}$ and $\mu_{pz}$. $\mu_{ii}$ is a decreasing function of $n$ hence $\mu$ starts decreasing when $n$ is increased beyond $1\times10^{13}$ cm$^{-3}$. At T=50K, $\mu_{ii}$ has the dominant contribution to the total mobility for $n \geq 1\times10^{15}$ cm$^{-3}$. $\mu_{po}$ is very high at T=50K indicating the fact that POP scattering is insignificant in low temperature. At T=300K, POP scattering is significant and has the dominant contribution in mobility. Figure 7 also indicates that at high temperature and high doping concentration $\mu_{pz}$ become less significant. We have observed similar trend for the TET phase (figure S5 and S6).

**Scattering Rates:** In order to have a better understanding of the observed transport properties we have analyzed scattering rates of different mechanisms at different temperature and carrier concentrations. Figure 8 shows the scattering rate vs. electron energy plots for the BCT configuration. We observe that all the scattering rates increases when the temperature rises. At low temp and low doping concentration piezoelectric scattering dominates in the low energy region (figure 8a). At T= 50K, the average electron energy is roughly $\frac{3}{2}KT = 0.0064$ eV. Hence piezoelectric scattering has been observed to have dominant contribution in the mobility at low temperature and low doping concentration. On the other hand POP energy is $\hbar\omega_{PO} = 0.04$ eV. At T=50K most of the electrons are in low energy region, making POP insignificant. At low temperature there is a predominant jump in the POP scattering rate. It is because of the fact that if the energy of an electron is less than 0.04 eV then it can be scattered only by absorption of an optical phonon. But, if electron energy is greater than 0.04 eV then scattering process involve both emission and absorption of polar optical phonons. POP scattering rate increases at T=300K and has the dominant contribution in the total mobility. Piezoelectric scattering rate also increases when temperature increases from 50K to 300K but POP scattering rate suppresses it, thereby making it insignificant at higher temp. When the temperature is 50K but the doping concentration increases, ionized impurity scattering rates is also found to increase and suppresses the contribution from piezoelectric scattering as is evident from figure 8b for $n = 1\times10^{17}$ cm$^{-3}$. The scattering rates of the TET configuration shows similar behaviour (Figure S7).

**IV. Conclusion:**

We have computed electron mobility of *n*-type AlGaAs$_2$ using Rode's iterative method with transport parameters calculated from DFT based approach. We have considered two different geometries of AlGaAs$_2$, viz. BCT-AlGaAs$_2$ and TET-AlGaAs$_2$. Both the TET and BCT phases are direct band gap semiconductors having PBE estimated band gap of 0.99 eV and 0.86 eV, respectively. Absence of mode with imaginary frequency in the phonon dispersions for both the phases confirms their stability. We have observed that the curvature of energy vs k-distance graph is higher for the BCT phase compared with the TET phase which results in higher electron mobility in the BCT phase than the TET phase. Therefore we can predict that the BCT phase will be more suitable for high mobility device applications as compared with the TET phase. We notice that piezoelectric scattering dominates in the low temperature and low doping concentration situation. At high doping concentration contribution from ionized impurity scattering significantly increases which suppresses the piezoelectric scattering contribution. Moreover we have also noted that at low temperature POP scattering is insignificant, however at room temperature POP scattering dominates.

Table I: Calculated material properties of AlGaAs$_2$ in both BCT and TET phases.

| Parameters | | AlGaAs$_2$ configurations | |
|---|---|---|---|
| | | BCT | TET |
| Primitive lattice vectors | $a$ (Å) | 7.03 | 4.06 |
| | $b$ (Å) | 7.03 | 4.06 |
| | $c$ (Å) | 7.03 | 5.74 |
| Angle between primitive lattice vectors | $\alpha$ (°) | 131.84 | 90 |
| | $\beta$ (°) | 131.84 | 90 |
| | $\gamma$ (°) | 70.48 | 90 |
| $\varepsilon_0$ | | 13.73 | 13.87 |
| $\varepsilon_\infty$ | | 11.56 | 11.71 |
| $E_D$ (eV) | | 15.25 | 14.2 |
| $E_g$ (eV) | | 0.99 | 0.86 |
| $\omega_{po}$ (THz) | | 10.4 | 10.70 |
| $c_l$ ($10^{10}$ N/m$^2$) | | 11.78 | 11.79 |
| $c_t$ ($10^{10}$ N/m$^2$) | | 4.04 | 4.05 |
| P | | 0.111 | 0.084 |
| $\rho$ (gm/cm$^3$) | | 4.33 | 4.33 |

$\varepsilon_0$ = low frequency dielectric constant, $\varepsilon_\infty$ = high frequency dielectric constant, $E_D$ = acoustic deformation potential, $E_g$ = electronic band gap, $\omega_{po}$ = Polar optical phonon frequency for the longitudinal mode, $c_l$ = longitudinal elastic constant, $c_t$ = transverse elastic constant, P = dimensionless piezoelectric coefficient, $\rho$ = density.

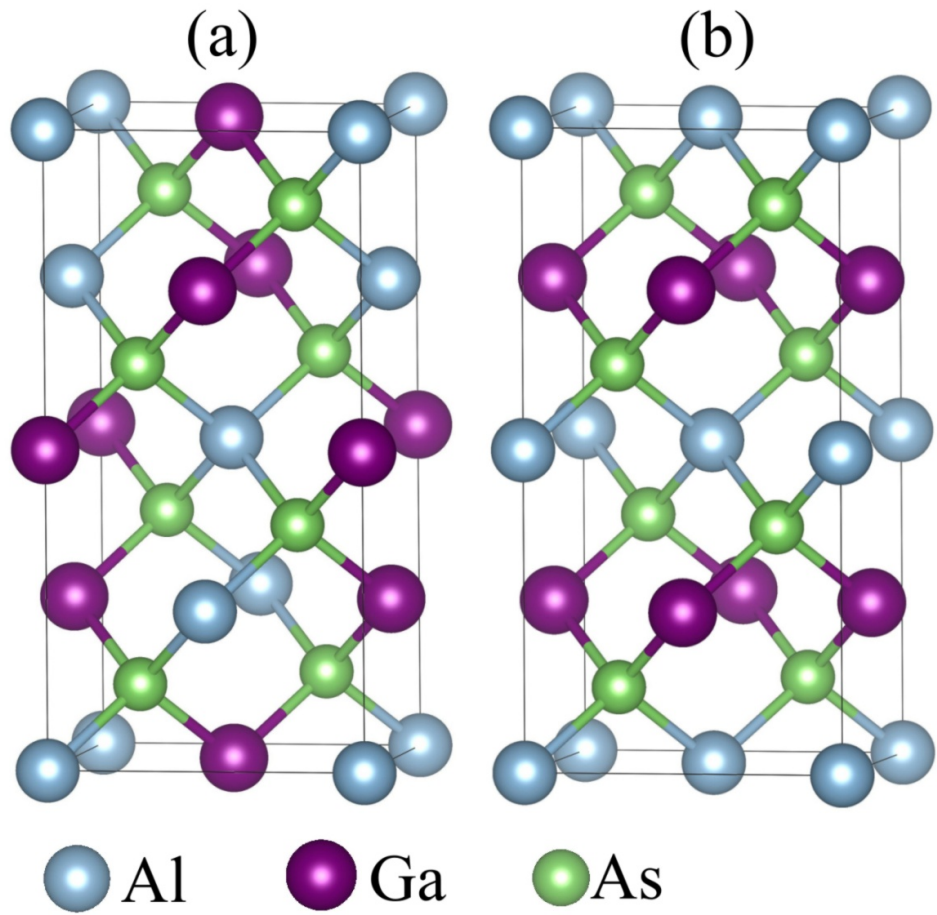

Figure 1: Conventional unit cell of AlGaAs$_2$: (a) BCT phase , (b) TET phase. Blue, purple and green spheres respectively corresponds to Al, GA and As atoms.

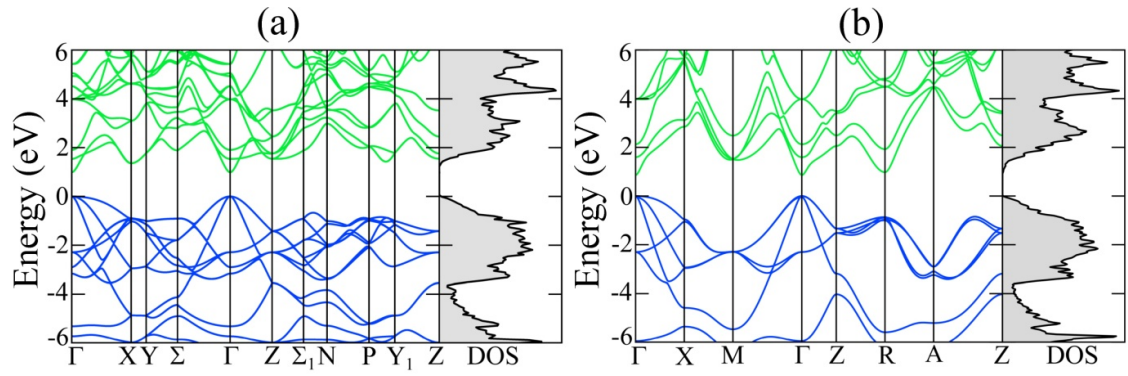

Figure 2: Band structure and density of states of AlGaAs$_2$: (a) BCT-AlGaAs$_2$ (b) TET-AlGaAs$_2$. Zero energy is set to valence band top. Density of states is in arbitrary unit.

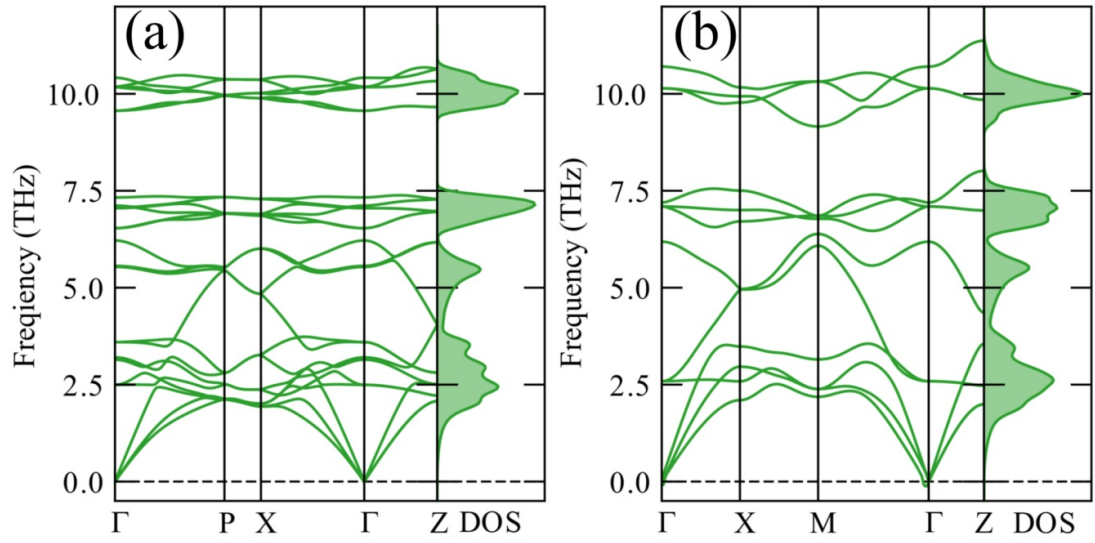

Figure 3: Phonon dispersion of (a) BCT-AlGaAs$_2$ and (b) TET-AlGaAs$_2$.

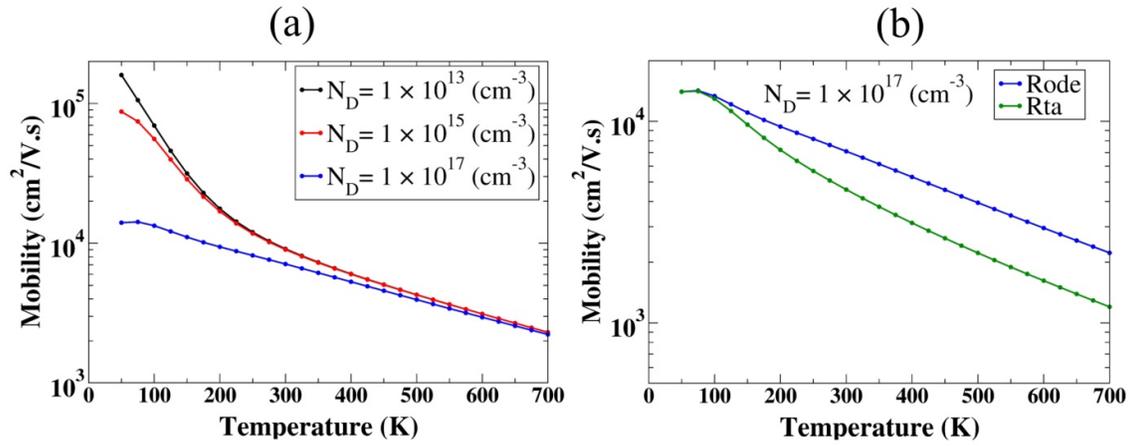

Figure 4: Variation of mobility with temperature for BCT-AlGaAs$_2$. (a) shows the mobility vs. temperature at different doping concentration with mobility computed using Rode's iterative method. The black, red and blue line indicates the corresponding plots for doping concentration $1\times10^{13}$ cm$^{-3}$, $1\times10^{15}$ cm$^{-3}$ and $1\times10^{17}$ cm$^{-3}$, respectively. (b) shows the comparison of mobilities calculated using Rode's method and Relaxation time approximation at the doping concentration $1\times10^{17}$ cm$^{-3}$. The blue line corresponds to the mobility estimated using Rode's method and the green line corresponds to RTA estimated mobility.

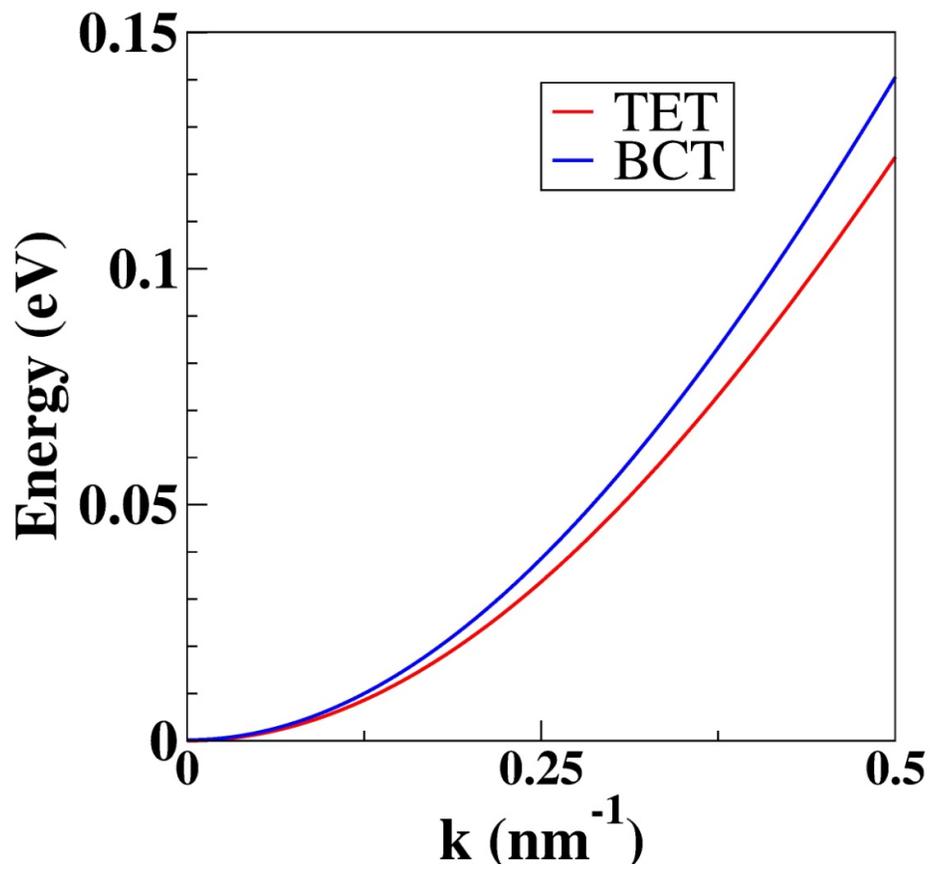

Figure 5: Energy of the electron in conduction band vs. k-distance plot. Blue and the red line are respectively the corresponding plots for the BCT and TET phases.

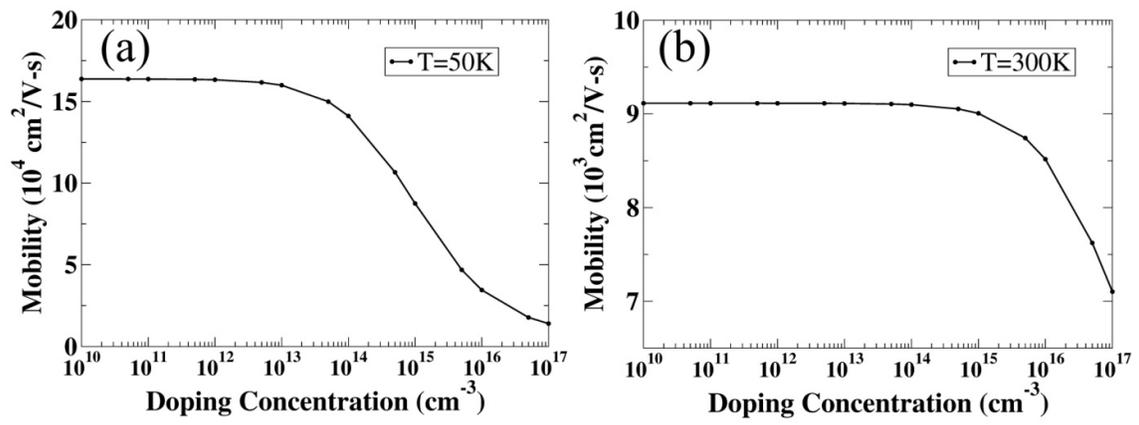

Figure 6: Variation of mobility with doping concentration at (a) T-50K and (b) T=300K for BCT-AlGaAs$_2$.

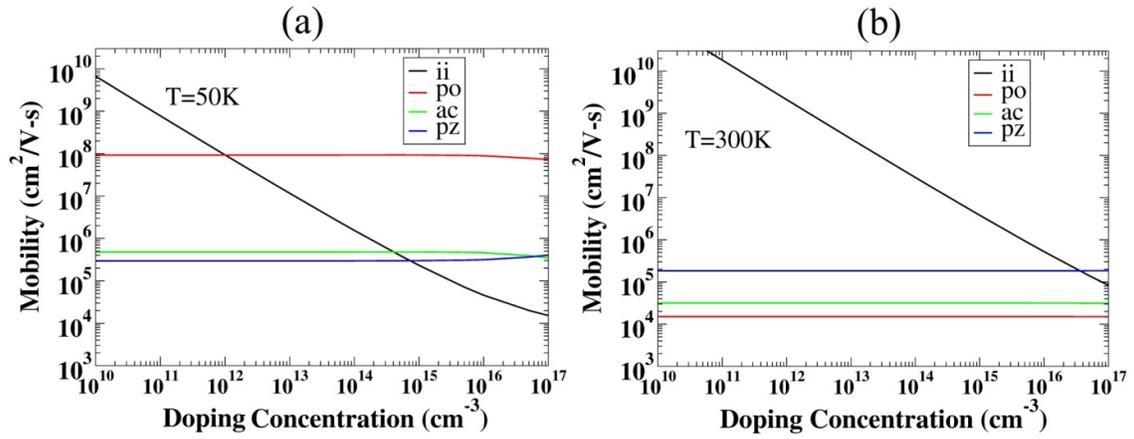

Figure 7: Contribution of different scattering mechanism to the mobility for BCT-AlGaAs$_2$ at (a) T=50K and (b) T=300K. The contributions from the ionized impurity, polar optical phonon, acoustic deformation potential and piezoelectric scatterings are indicated by the black, red, green, and blue lines, respectively.

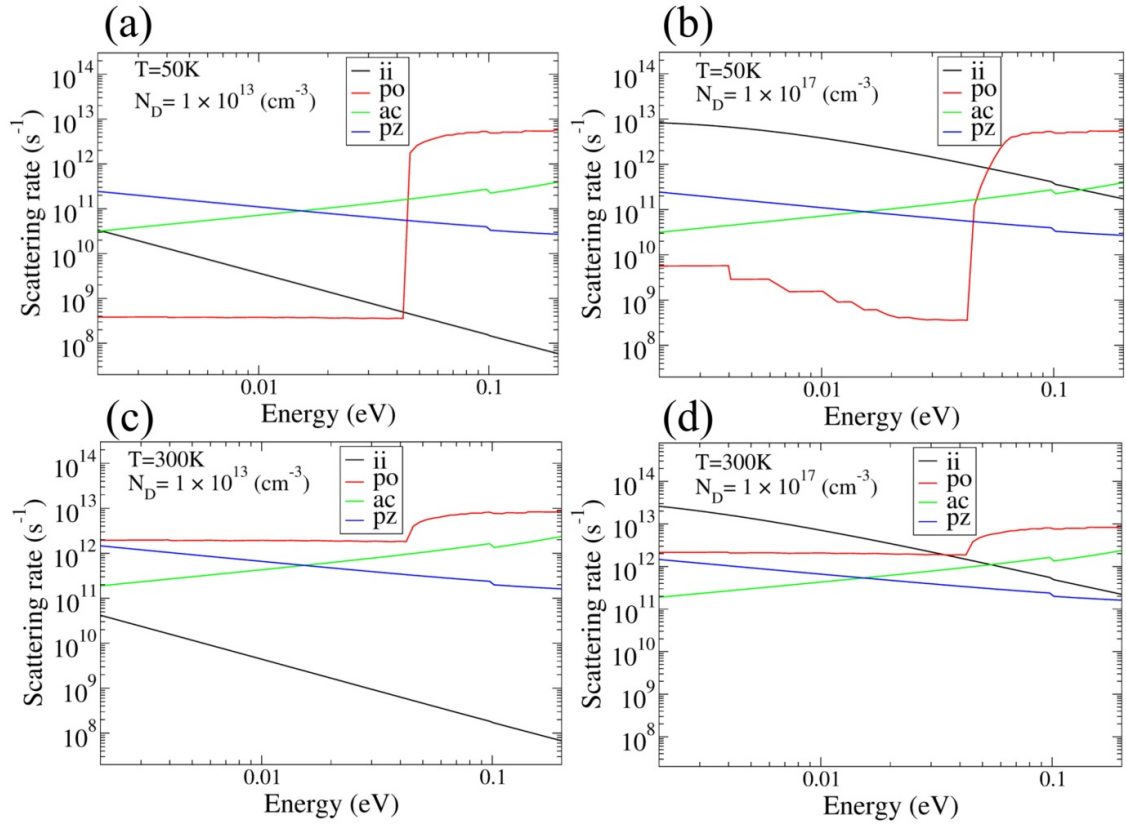

Figure 8: Scattering vs. electron energy plot for BCT-AlGaAs$_2$ at (a) T=50K and doping concentration=$1\times10^{13}$ cm$^{-3}$ (b) T=50K and doping concentration=$1\times10^{17}$ cm$^{-3}$ (c) T=300K and doping concentration=$1\times10^{13}$ cm$^{-3}$ (d) T=300K and doping concentration=$1\times10^{17}$ cm$^{-3}$.

# Supplementary Material

# Semi-classical electronic transport properties of ternary compound AlGaAs$_2$: Role of different scattering mechanisms.

Soubhik Chakrabarty[a], Anup Kumar Mandia[b], Bhaskaran Muralidharan[b], Seung Cheol Lee[a,*], and Satadeep Bhattacharjee[a,*]

[a]Indo-Korea Science and Technology Center, Bangalore 560065, India

[b]Indian Institute of Technology, Mumbai-400076, India

*Corresponding Author E-mail: seungcheol.lee@ikst.res.in, satadeep.bhattacharjee@ikst.res.in

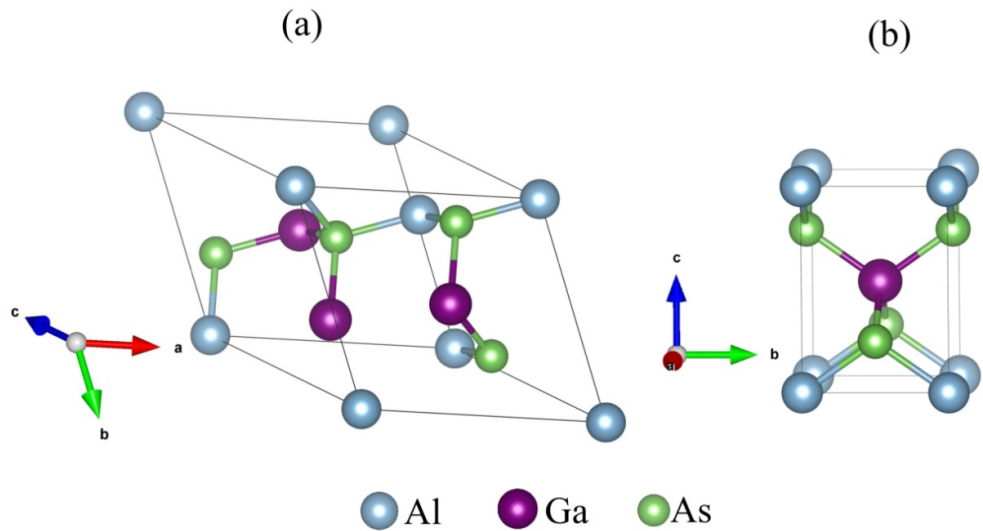

Figure S1: Primitive unit cell of AlGaAs$_2$: (a) BCT phase, (b) TET phase. Blue, purple and green spheres respectively corresponds to Al, GA and As atoms.

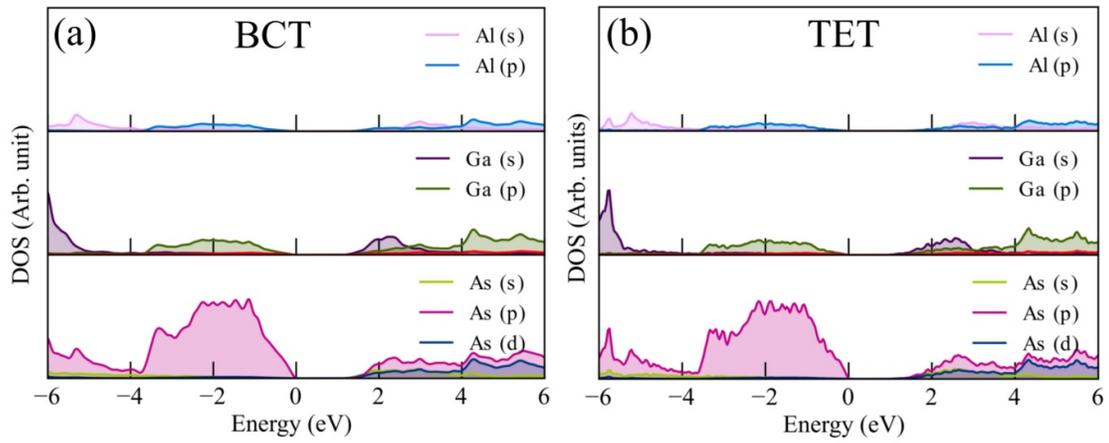

Figure S2: Atom and orbital projected density of state of AlGaAs$_2$: (a) BCT-AlGaAs$_2$ (b) TET- AlGaAs$_2$. Zero energy is set to valence band top. Density of states is in arbitrary unit.

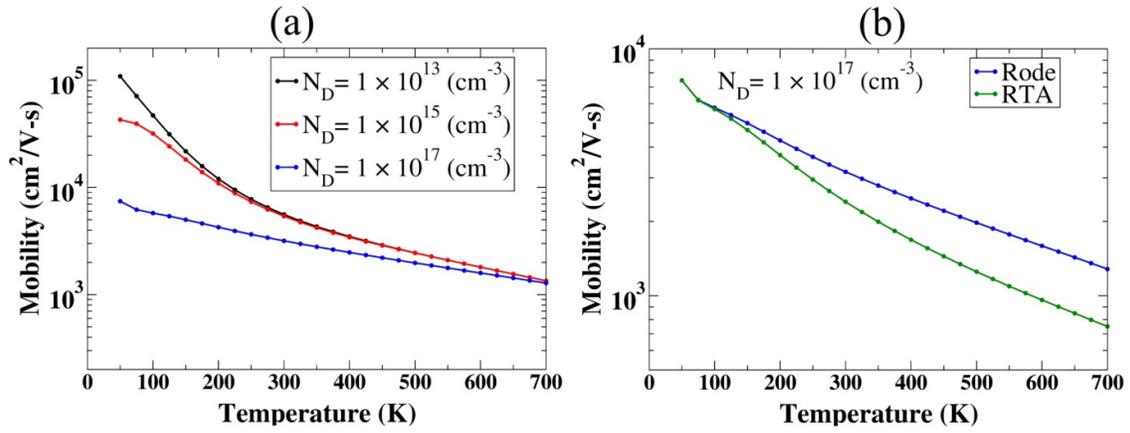

Figure S3: Variation of mobility with temperature for TET-AlGaAs$_2$. (a) shows the mobility vs. temperature at different doping concentration with mobility computed using Rode's iterative method. The black, red and blue line indicates the corresponding plots for doping concentration $1\times10^{13}$ cm$^{-3}$, $1\times10^{15}$ cm$^{-3}$ and $1\times10^{17}$ cm$^{-3}$, respectively. (b) shows the comparison of mobilities calculated using Rode's method and Relaxation time approximation at the doping concentration $1\times10^{17}$ cm$^{-3}$. The blue line corresponds to the mobility estimated using Rode's method and the green line corresponds to RTA estimated mobility.

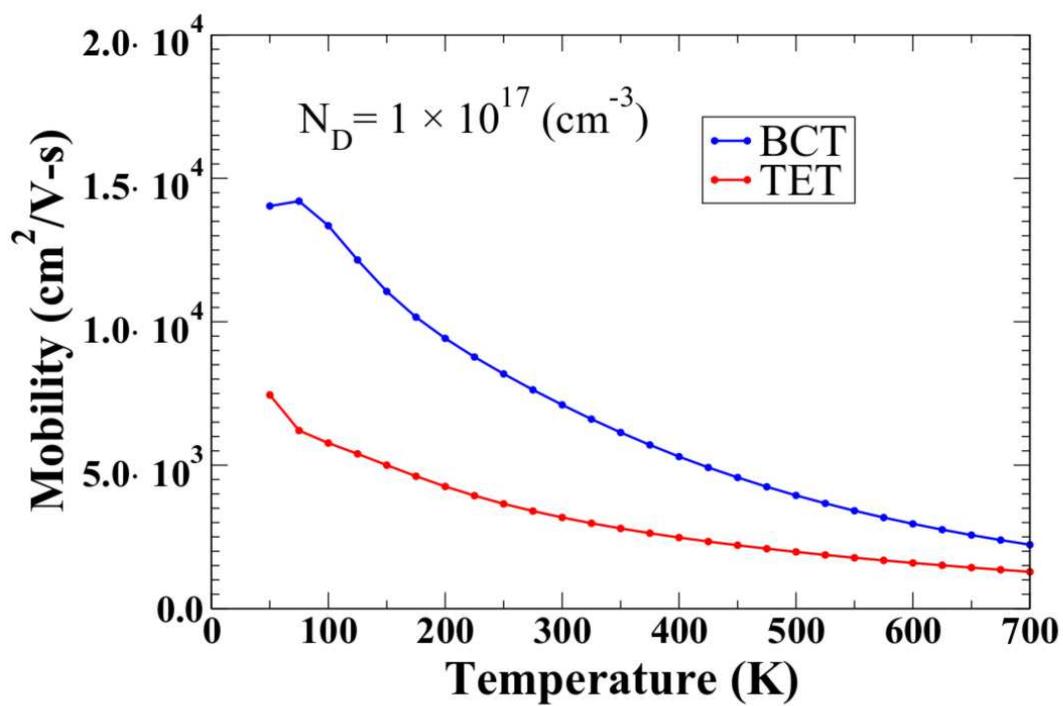

Figure S4: Comparison of mobilities of BCT-AlGaAs$_2$ and TET-AlGaAs$_2$.

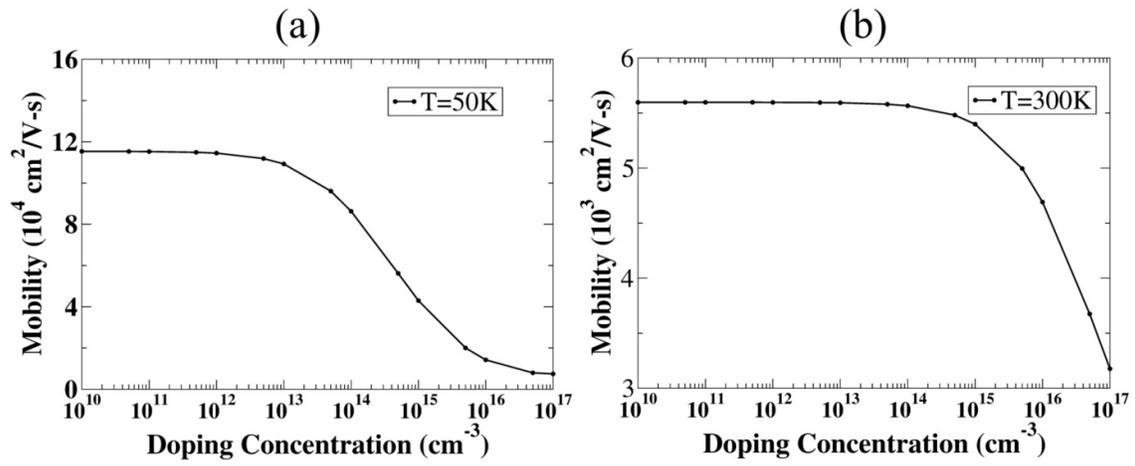

Figure S5: Variation of mobility with doping concentration at (a) T-50K and (b) T=300K for TET-AlGaAs$_2$.

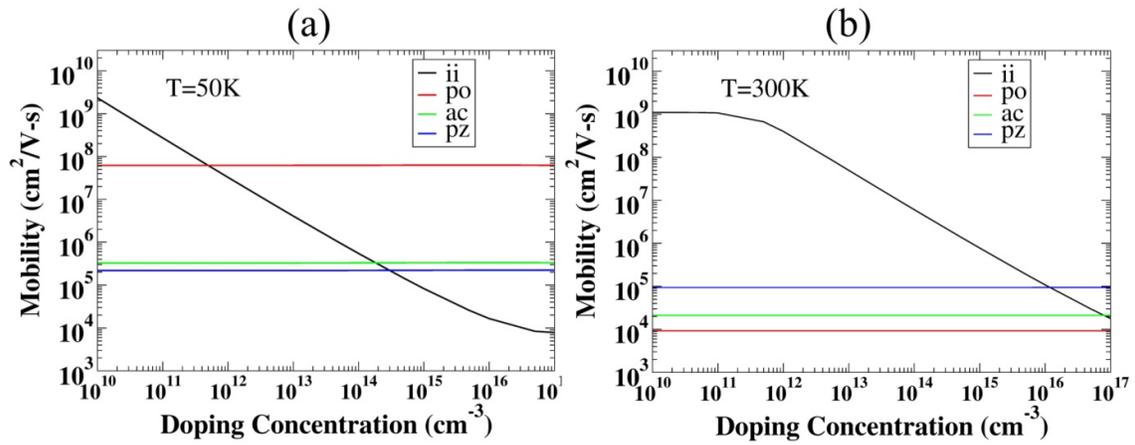

Figure S6: Contribution of different scattering mechanism to the mobility for TET-AlGaAs$_2$ at (a) T=50K and (b) T=300K. The contributions from the ionized impurity, polar optical phonon, acoustic deformation potential and piezoelectric scatterings are indicated by the black, red, green, and blue lines, respectively.

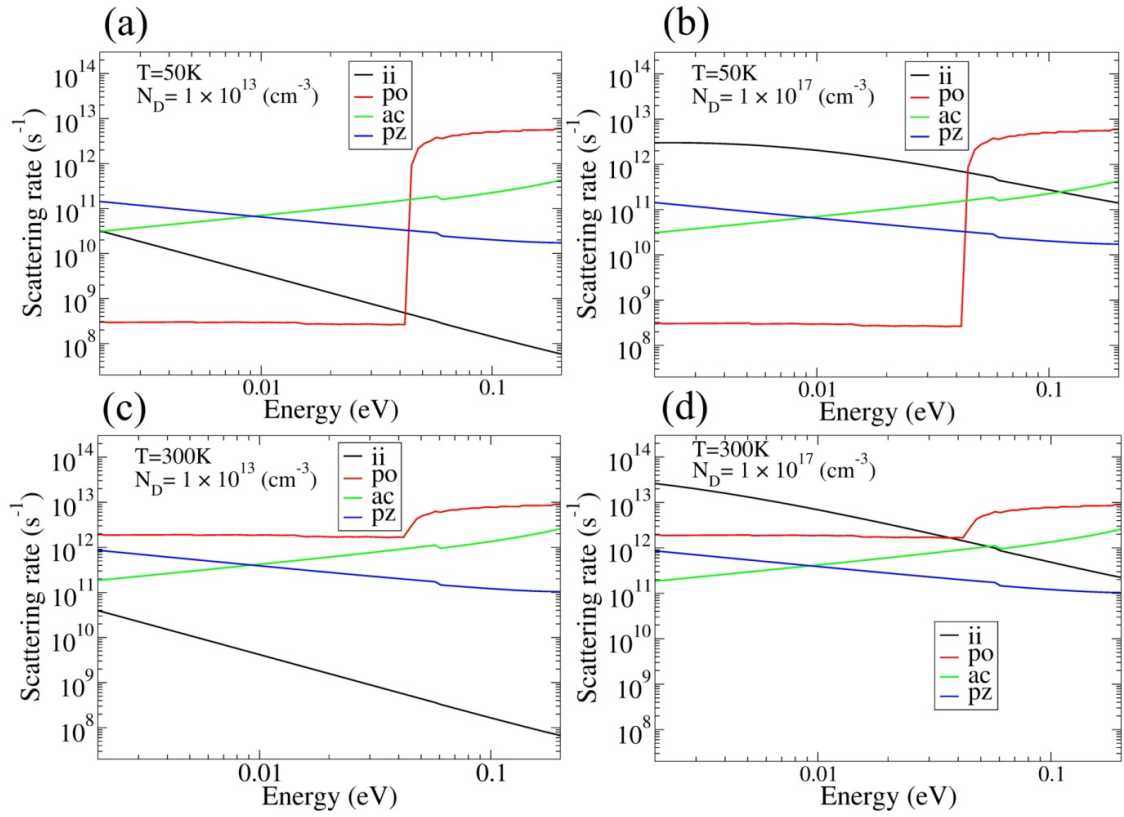

Figure S7: Scattering vs. electron energy plot for TET-AlGaAs$_2$ at (a) T=50K and doping concentration=$1\times10^{13}$ cm$^{-3}$ (b) T=50K and doping concentration=$1\times10^{17}$ cm$^{-3}$ (c) T=300K and doping concentration=$1\times10^{13}$ cm$^{-3}$ (d) T=300K and doping concentration=$1\times10^{17}$ cm$^{-3}$.